\documentclass[preprints,article,accept,moreauthors,pdftex]{Definitions/mdpi}
\usepackage{isotope}
\graphicspath{{figures/}}
\firstpage{1} 
\pubvolume{1}
\issuenum{1}
\articlenumber{0}
\pubyear{2023}
\copyrightyear{2023}
\datereceived{12 December 2022}
\daterevised{19 January 2023} 
\dateaccepted{31 January 2023} 
\datepublished{} 
\hreflink{https://doi.org/10.3390/s23041945} % If needed use \linebreak

\pdfoutput=1

\Title{Ambient Dose and Dose Rate Measurement in SNOLAB Underground Laboratory at Sudbury, Ontario, Canada}

\TitleCitation{Ambient Dose and Dose Rate Measurement in SNOLAB Underground Laboratory at Sudbury, Ontario, Canada}

\Author{{Victor V. Golovko}
 $^{1,}$*\orcidA{}, Oleg Kamaev  $^{1}$\orcidB{}, Jiansheng Sun $^{1}$, Chris J. Jillings $^{2}$, Pierre Gorel $^{2}$ \mbox{and Eric~V$ \acute{\text a} $zquez-J$ \acute{\text a} $uregui $^{3}$}}

\AuthorNames{Victor V. Golovko, Oleg Kamaev, Chris J. Jillings, Pierre Gorel and Eric~V$ \acute{\text a} $zquez-J$ \acute{\text a} $uregui}

\AuthorCitation{Golovko, V.V.; Kamaev, O.; Sun, J.; Jillings, C.J.;  Gorel, P.; Eric, V.-J.}
\address{%
$^{1}$\quad {Canadian Nuclear Laboratories,} 286 Plant Road, Chalk River, ON K0J~1J0, Canada; oleg.kamaev@cnl.ca (O.K.); jiansheng.sun@cnl.ca (J.S.)\\
$^{2}$\quad {SNOLAB,} 
 Lively, ON P3Y 1N2, Canada; Chris.Jillings@snolab.ca (C.J.J.); Pierre.Gorel@snolab.ca (P.G.)\\
$^{3}$\quad Instituto de F\'{\i}sica, Universidad Nacional Aut$ \acute{\text o} $noma de M$ \acute{\text e} $xico, A. P. 20-364, Mexico City 01000, 
 Mexico; ericvj@fisica.unam.mx \\
}

\corres{Correspondence: victor.golovko@cnl.ca}

\abstract{The paper describes a system and experimental procedure that use integrating passive detectors, such as  thermoluminescent dosimeters (TLDs), for the measurement of ultra-low-level ambient dose equivalent rate values at the underground SNOLAB facility located in Sudbury, Ontario, Canada. Because these detectors are passive and can be exposed for relatively long periods of time, they can provide better sensitivity for measuring ultra-low activity levels. The final characterization of ultra-low-level ambient dose around water shielding for ongoing direct dark matter search experiments in Cube Hall at SNOLAB underground laboratory is given. %The comparison with the measurement results obtained using the gamma-ray spectrometry method validates the results reported in this paper. 
The conclusion is that TLDs provide reliable results in the measurement of the ultra-low-level environmental radiation background.}

%; Monte Carlo simulation} 

\begin{document}

%%%%%%%%%%%%%%%%%%%%%%%%%%%%%%%%%%%%%%%%%%
%\setcounter{section}{-1} %% Remove this when starting to work on the template.

% The order of the section titles is: Introduction, Materials and Methods, Results, Discussion, Conclusions for these journals: aerospace,algorithms,antibodies,antioxidants,atmosphere,axioms,biomedicines,carbon,crystals,designs,diagnostics,environments,fermentation,fluids,forests,fractalfract,informatics,information,inventions,jfmk,jrfm,lubricants,neonatalscreening,neuroglia,particles,pharmaceutics,polymers,processes,technologies,viruses,vision

\section{Introduction}

The background models for ordinary particles and their contributions in direct detection experiments, which search for dark matter particles interacting with ordinary matter in a terrestrial detector target, are extremely important. These background models are used in various ongoing and future dark matter search experiments (see, for example,~\cite{angloher2012results,tan2016dark,akerib2017results,Aalbers_2016,amaudruz2018first,ajaj2019electromagnetic,collaboration2019xenon1t,Agnes_2021,aalbers2022next}) and are studied, in detail, for their contributions in the signal search region using extensive Monte Carlo simulations. To~reduce background contributions from ordinary particles, typically the direct dark matter search experiments are placed deep in underground laboratories, where contributions from ordinary background sources, such as cosmic rays, are greatly reduced. However, the environmental background still exists and needs to be known or~measured.

\color{black}Underground facilities in North America nowadays represent  suitable places to perform numerous experiments that are highly sensitive to any source of background radiation. Reviews of the major low-level background underground facilities in North America that are currently available for ``state-of-the-art'' physics experiments are provided by~\cite{henning2011north,lesko2015north}. The~deep underground facilities are required by the science community due to rigorous radiation background constraints to afford shielding from penetrating cosmic rays and their induced by-products~\cite{hall2020snolab}. Many modern research topics require laboratory environments as free as possible from cosmic radiation and radioactive isotopes. A~dedicated effort to determine how to comply with such requirements and, at the same time, to provide a comprehensive list of current and potential underground experiments has been made recently for SNOLAB in~\cite{smith2012snolab,duncan2010construction} and the Sanford underground research facility at Homestake in~\cite{lesko2012sanford,lesko2015sanford}. 

SNOLAB is located near Sudbury, Ontario, at the operating Creighton nickel mine (owned by Vale Inco, now Vale Canada Limited) 
and has vertical access~\cite{duncan2010construction,smith2012snolab}. It is an extension of the existing facilities that were constructed originally for the Sudbury Neutrino Observatory (SNO) solar neutrino experiment~\cite{duncan2010construction}. The~entire lab, which hosts many experiments, is operated as a large clean~room.

\begin{figure}[t]
	\includegraphics[width=\textwidth]{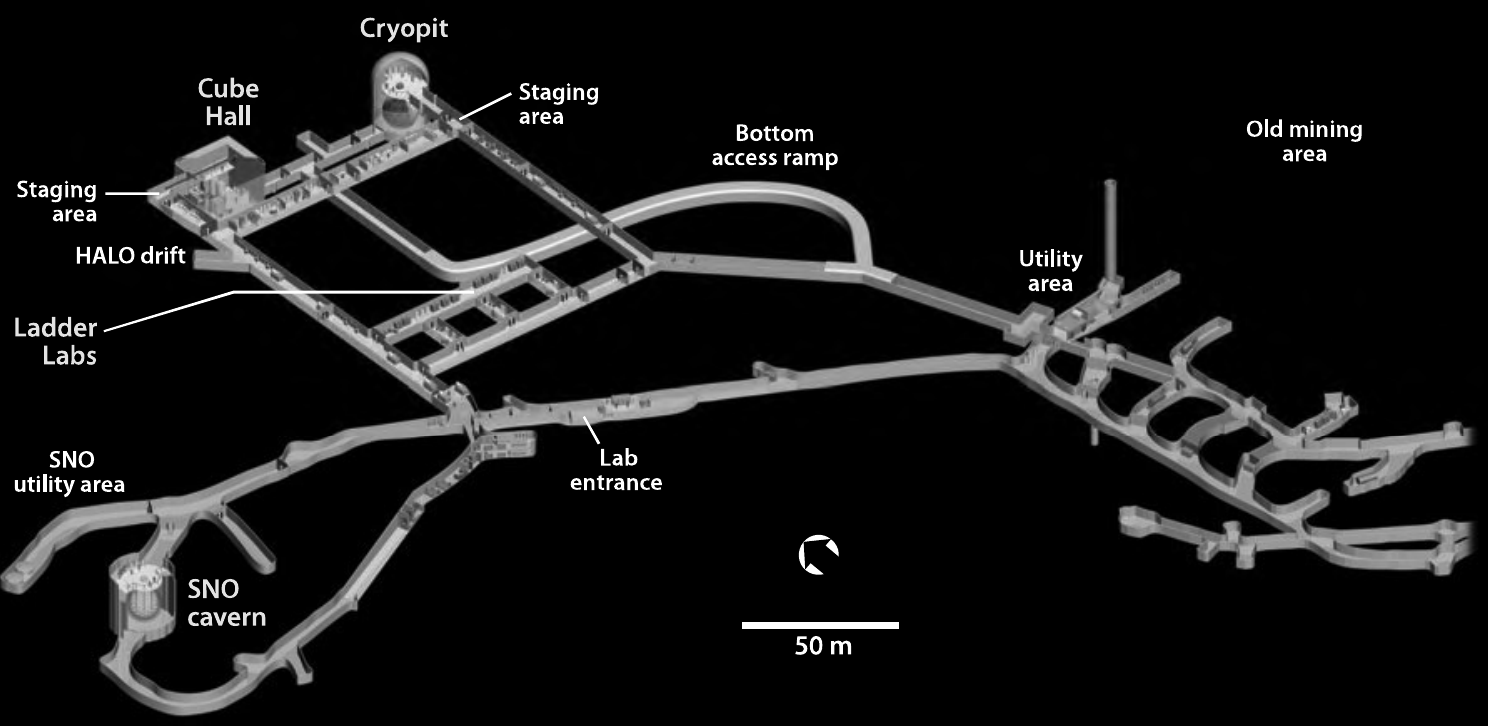}	
	\caption{{A 3D model of the SNOLAB in Sudbury,} 
		Ontario, Canada (image credit: \cite{duncan2010construction}). Scale and direction to the geographical north are also~provided.}
	\label{fig:SNOLAB3D}
\end{figure}

\begin{figure}[t]
	\centerline{
		\centering
		\includegraphics[width=\textwidth]{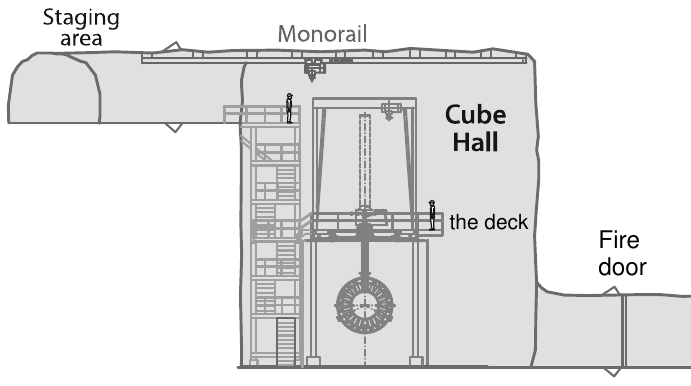} 
		%\includegraphics[width=\textwidth, trim=15 10 260 10, clip=true]{p6gr-crop} 
		% left bottom right top
	}
	\caption{Cross-section of the Cube Hall (image credit: \cite{duncan2010construction}). The~integrating passive detectors were placed around the water shielding (or water tank) of the DEAP-3600 experiment (facing northward). The design and construction features of the DEAP-3600 detector are given in~\cite{AMAUDRUZ20191}.}
	\label{fig:CubeHall}
\end{figure}

SNOLAB has a rather complicated layout (see Figure~\ref{fig:SNOLAB3D}). It consists of an ``old'' part that hosted the original SNO~\cite{boger2000sudbury} experiment (now transformed to SNO+; see the latest development in~\cite{Caden_2020}) and a relatively ``new'' part that has been excavated to host  many experiments, and that shares the same lab infrastructure. It is currently the deepest underground laboratory in North America, with a 2~km overburden (estimated to be equivalent to 6000~m of water). It is surrounded by norite granite rock. SNOLAB currently houses several experiments, including the dark matter experiment using argon pulseshape discrimination (DEAP-3600), currently the world’s largest and most sensitive liquid argon dark matter detector, which is located in Cube Hall (see Figure~\ref{fig:SNOLAB3D} and cross-section of Cube Hall with DEAP-3600 experiment in Figure~\ref{fig:CubeHall}). The~Cube Hall clean room space is approximately 18.3~m in length, 15~m in width, and~19.7~m in height and includes the experimental areas hosting MiniCLEAN~\cite{akashi2019triplet} and DEAP-3600~\cite{AMAUDRUZ20191} experiments, as well as ancillary spaces. DEAP-3600 and MiniCLEAN are searching for interactions of weakly interacting massive particles (WIMPs) with argon nuclei. WIMPs are currently the best candidates to explain dark matter, which makes up roughly 85\% of the mass content of the~universe. {{At} the time of the ambient dose measurement with integrating passive detectors, the~MiniCLEAN experiment was in the decommissioning state. Currently, the Cube Hall clean room is also occupied by the light dark matter particle candidate search NEWS-G (New Experiments With Spheres-Gas) experiment~\cite{BALOGH2021164844}}.

There are various ways to characterize the low-level gamma-ray background at  underground laboratory facilities. For~example, a~low-background high-purity germanium (HPGe) gamma spectrometer has been used to measure environmental radioactivity (i.e.,~the natural radioactivity) in the underground Callio Lab facility~\cite{gostilo2020characterisation}. Specific to Cube Hall at SNOLAB, some techniques used to characterize the electromagnetic background at the DEAP-3600 detector are discussed in detail in~\cite{ajaj2019electromagnetic}.  DEAP-3600 has exquisite control of radioactive backgrounds from the argon, detector materials, and~surrounding rock. The~detector (originally designed to host 3600~kg) consists of a 3.2-tonne quantity of liquid argon inside a radiopure acrylic vessel. Attached to the acrylic vessel are 255 photomultiplier tubes (PMTs) 8'' in diameter, which are sensitive enough to detect single photons. The~PMTs are separated by 50-cm-long 8'' diameter acrylic light guides~\cite{AMAUDRUZ20191}. The~detector is placed inside an 8-m-tall water tank {(see Figure}~\ref{fig:CubeHall}) to shield it from radioactivity in the rock and to tag cosmic ray~muons.

 The Physikalisch-Technische Bundesanstalt established an underground laboratory for dosimetry and spectrometry  at the Asse salt mine, near Braunschweig, where a coaxial low-background HPGe detector with extended shielding was exploited~\cite{neumaier2000ptb}. A detailed investigation of the dose rate within the underground low-level background laboratories at the Unirea salt mine (Slanic-Prahova) has shown a relatively uniform distribution varying between 1.3$\pm$0.3~nSv~h$^{-1}$ outside  the Low-Level Background Laboratory (LLBL) and 1.6$\pm$0.3 nSv h$^{-1}$ inside  it~\cite{margineanu2008slanic,margineanu2009external}. This measurement was performed by means of a calibrated scintillator rate meter. Two rounds of measurement of ambient dose rate were obtained at a significant number of points situated inside the Praid salt mine, by using the very sensitive Romanian thermoluminescent dosimetry (TLD) system, as reported in~\cite{stochioiu2012area}. Authors in~\cite{stochioiu2012area} showed a consistency between gamma-ray spectrometry results and results from the TLD-based system.

%\unskip

It is well known that the TLD systems have the advantage of being cheap, compact, easy to manipulate and~allow the monitoring of various places in an extended location, such as Cube Hall. The~continuous survey of the areas of interest, without~disturbing the experiments can be assured by using TLD-based systems. Thus, one registers every variation in the background values that could influence the basic measurement results. Measurements have been performed around the water shielding for the DEAP-3600 experiment~\cite{amaudruz2018first,AMAUDRUZ20191}. As~can be seen from the report on natural gamma-ray backgrounds in the underground LLBL (\cite{margineanu2009external,gostilo2020characterisation} or~\cite{neumaier2000ptb}), the~rate could be quite uniform and stable all over the lab, or it could vary (see, for example,~\cite{stochioiu2012area}). 

Many factors could contribute to either  uniform or non-uniform distributions of natural environmental gamma-ray backgrounds in underground labs. In~this paper, we apply a measurement technique using many integrating passive detectors, such as TLDs, that are used successfully at Chalk River Laboratories (CRL) to monitor doses. The~methods and approach that we used are subject to a regular annual independent blind test from the Canadian regulator that CRL Dosimetry Services has successfully~passed. 

%The water shielding for DEAP-3600~\cite{amaudruz2018first}  detector has been designed based on extensive Monte Carlo studies for shielding optimization for background rejections~\cite{AMAUDRUZ20191} that also protected the detector from environmental gamma background. 
The DEAP-3600 detector searches for direct interaction with dark matter particles using noble gas. Using extensive Monte Carlo studies, optimization of the water shielding has been performed for the muon background rejection and the protection against environmental gamma background. These studies assume a uniform distribution of environmental gamma flux background around a large-scale detector, such as DEAP-3600. This approach allows us to get a very conservative upper limit. As~we mentioned previously, the~natural gamma-ray background around the detector is assumed to be uniform. Integrating passive TLD detectors is the tool of choice for testing this assumption.
%In this paper, we report the external dose measurements that were done underground at SNOLAB in Cube Hall placing TLD dosimeters around the water shielding of DEAP-3600 experiment.

\subsection*{Scope of~Work}
\label{sow}

For this project, we have identified the following~objectives:
\begin{itemize}
	\item	First, would it be possible to use integrating passive detectors, such as TLDs, for ambient dose and dose rate measurements in low-level background underground facilities, in~particular at SNOLAB? To the best of our knowledge, SNOLAB is one of the world's deepest operating underground facilities~\cite{doi:10.1146/annurev-nucl-102115-044842}. The~underground laboratory was created to provide an almost background-free environment for very sensitive experiments, such as the direct observation of potential dark matter particle candidates. One of the current challenges was that   SNOLAB  hosts an operational DEAP-3600 detector~\cite{adhikari2022first}, which, on a regular basis, takes calibration measurements using quite strong external sources; therefore, it was a logistical challenge to find and organize a four-week-long interval for TLD deployment.  
	\item	Second, in~the ongoing experiment, the performance of the detector shielding  against various types of background radiations is studied via careful and detailed Monte Carlo simulations. One of the assumptions in these studies is the distribution of background radiation, such as gamma radiation, around the detector shielding. A~typical assumption is that at the detector shielding surface, the radiation distribution is uniform. Here, we would like to test this assumption.
	%\item	Thirdly, if one could measure the ambient dose and dose rate measurements at the surface of the detector shielding using integrated passive detectors would it be possible to simulate those responses using simplified Monte Carlo model.
\end{itemize}

For next-generation ultra-sensitive experiments for the direct detection of dark matter using a multi-ton noble liquids technology, such as DarkSide~\cite{Agnes_2021} and DARWIN~\cite{Aalbers_2016,aalbers2022next}; the knowledge of the external sources of backgrounds is ultimately important. Typically, the~design of the shielding from external sources of backgrounds for such experiments is made based on conservative and independent estimates for the upper limit of environmental gammas assuming uniform distributions. With~this work, we would like to show that integrating passive detectors, such as TLD, are capable of measuring ambient dose and dose rate around dark matter experiments. Proposed next-generation  direct-detection dark matter experiments will be located in the ultra-low-level ambient doses surroundings. The~deployment of the integrating passive detectors, such as TLDs, is recommended at the beginning and at the end of the operational campaign for the dark matter detector with an exposure period of around three months or more (see Section~\ref{sec:Dis}) to reach a reasonable uncertainty for ambient dose and dose rate~measurements.

\color{black}
%%%%%%%%%%%%%%%%%%%%%%%%%%%%%%%%%%%%%%%%%%
\section{Materials and~Methods}

To obtain a cumulative environmental dose measurement of the direct  gamma radiation, integration dosimeters, such as TLDs, are used. A~typical CRL dosimeter system has two thermoluminescent elements to measure radiation exposure, one of which uses an ``open'' window  to minimize attenuation of beta radiation; the second incorporates an aluminum filter to attenuate beta radiation and, at the same time,  minimize the attenuation of low-energy photon~radiation. 

\begin{figure}[t]
	\centerline{
		\centering
		\includegraphics[width=0.99 \columnwidth ]{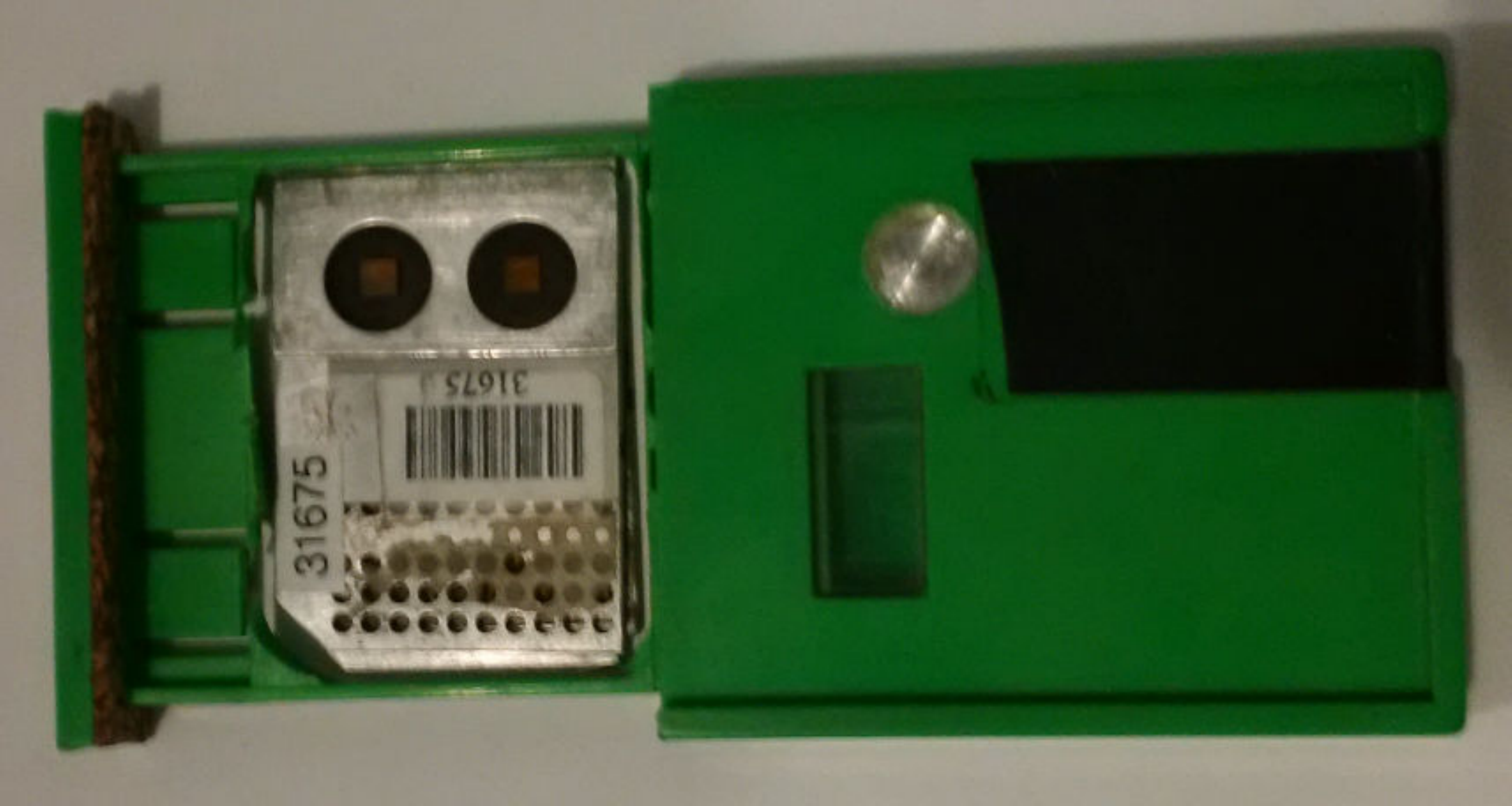} 
	}
	\caption{TLD housing (badge) with ``thin'' (TLD-700) and ``thick'' (TLD-100) chips  were used in this study. To~improve statistics, two aluminum plaques with TLD chips were used, one on top of the other. Note that two aluminum plaques, ``front'' and ``rear'', are located in the badge on top of each~other.}
	\label{fig:badge}
\end{figure}

All environmental photon doses in this work were measured using integrating passive detectors, such as Harshaw LiF:Mg,Ti chips, which have natural isotopic abundances of \isotope[6]{Li} and \isotope[7]{Li}. This material is widely known as TLD-100. The~chip dimensions are 3.2~ $\times$ 3.2~ $\times$ 0.89~mm. This TLD chip is referred to as a ``thick'' chip (see Figure~\ref{fig:badge}).

Figure~\ref{fig:badge} also shows integrating passive detectors that are used to measure doses from photon and beta radiation. These doses are measured using Harshaw LiF:Mg,Ti chips enriched in the isotope \isotope[7]{Li} (i.e.,~99.99\% \isotope[7]{Li}). This material is widely known as TLD-700. The~chip dimensions are 3.2  $\times$ 3.2  $\times$ 0.38 mm. This TLD chip is referred to as a ``thin'' chip. {As we were interested in the environmental photon doses, the~dose readings from a ``thin'' chip were not used in this work}. 

Before being mounted on dosimeter plaques, all TLD chips undergo a process of sensitization, whereby a very large dose is delivered to the chips in a gamma cell, after which they are annealed at elevated temperatures in the presence of ultraviolet light. This results in a quasi-permanent increase in sensitivity by a factor of nominally three. The~TLD chips are mounted on Kapton tape, which is glued onto a stamped aluminum plaque, the~chips occupying cavity positions in the~plaque.

Dosimeter plaques, when issued for use, are inserted into badges, which consist of a two-piece plastic molding with appropriate recesses and cavities for the dosimeters. The~badge casing contains 1.85-mm-thick aluminum filters positioned in the front and back such that the thick TLD chip is positioned between them in the assembled dosimeter badge. The~badge casing also has two cavities, front and back, such that the ``thin'' TLD is not filtered by the badge casing. To~protect the dosimeters from dust and light, opaque tape with a mass thickness of nominally  7~mg~cm$^{-2}$ ($ \simeq $  60~$\mu$m) is wrapped around the badge casing to cover the front and back~cavities.

The front of the badge casing contains an identification card with a mass thickness of nominally  0.107~g~cm$^{-2}$, consisting of 80\% polyvinyl chloride  and 20\% polyethylene terephthalate. Cut-outs in this card are positioned such that the plaque identification (ID) is still visible and~such that no additional filtration is introduced in front of the thin chip. Figure~\ref{fig:badge} shows one of the badges used in this work. It has been opened to reveal its internal~structure.

The aluminum filter in the front and back of the ``thick'' TLD chip also provides a fundamental function, whereby the primary radiation field that consists of indirectly ionizing radiation (for example, gamma rays) that passes through the filter medium may generate  secondary charged particles. The~secondary charged particles reach charge-particle equilibrium (CPE) within the sensitive volume of the TLD chip, assuming that the primary radiation uniformly irradiates the entire assembly (aluminum filter and TLD chip) and is negligibly attenuated in traversing it. The~TL mechanism under CPE is described in detail elsewhere~\cite{attix1975further}.

TLD plaques are read on Harshaw 6600 automatic TLD readers. For~ simplicity,  in the text, we refer to them as ``reader(s)''. The~TLD monitoring systems for CRL use calibration irradiation facilities that are traceable to Canadian national standards through the free-in-air exposure quantity. The~dosimetry services at the CRL site calibrate their thick and thin-dose TLD monitoring systems to report personal~dose equivalent.

Sources for photon irradiation energy and angle type-testing  consisted of \isotope[60]{Co} gamma-ray sources from the Health Physics Irradiation Facility at CRL. {Requirements} for energy and angle type-testing for the CRL dosimetry system used in this work are outlined in a regulatory document~\cite{canadian2018document}.
The Health Physics Irradiation Facility photon irradiations are traceable through measurements using ion chambers and electrometers with a calibration that is traceable to national reference~standards.

For calibration irradiations, the~response by several influencing quantities are expressed relative to those under reference conditions, or~else the variability is type-tested without regard to absolute dose.
For these type-tests, it is sufficient that the irradiation geometry is reproducible and that the exposure time is accurately~known.

In general, integrating passive detectors, such as TLD chips, which are made by the same manufacturing process, have the same amount of TL materials on the elements, but~since this amount is a very small quantity, the~amount of TL materials on each element can vary slightly. Thus, the~element response of the TLD changes with the variation in the amount of TL materials. Furthermore, the~element response of the TLD is gradually decreased by physical conditions, such as impact, heat, humidity, etc., due to repeated use. It is, therefore, necessary to check and compensate for the element response periodically to obtain more accurate and reliable TLD readouts for dose~assessment.

The calibration of all the TLD was performed at CRL, where the gamma irradiators are controlled under a rigorous quality assurance program. The~photon fields used in this work were gamma rays from \isotope[60]{Co} that were produced following the methods of ISO~4037-1~\cite{Standardization2019doc}. Gamma-ray irradiations for TLD calibration were performed using the GC60 Gamma Beam Irradiator from Hopewell Designs located at~CRL.

By using a set of calibration dosimeters and a gamma-ray \isotope[60]{Co} irradiation source, the~reader's performance can be kept at a constant level in spite of high-voltage changes, repairs, dirt accumulation, or~long-term drift. {At} CRL, typically, a set of six calibration dosimeters are used that are exposed to a known photon dose.  The~calibration factor for the readers is known as the reader calibration factor or~$RCF$. This factor converts the raw charge data from the PMTs (in nanocoulombs) to dosimetric units (rems, for~example) for input to the following formula: 

\begin{linenomath*}
\begin{equation}
	{H_i} = \frac{ Q_i - Q_0} { RCF },
	\label{Eq:app:RCF}
\end{equation}
\end{linenomath*}
where $H_i$ is the measured dose in detector $i$, $Q_i$ is the number of counts in detector $i$, $RCF$ is the reader calibration factor, and~$ Q_0 $ is the average of the count of the zero-dose readings (or background readings). {Note} that the same formula could be used to deduce $RCF$ for the reader.

The uncertainty of the $ RCF $ could be estimated based on the quality assurance and quality control program at CRL Dosimetry Services, which requires annual independent tests for ``thick'' TLD chips as outlined in previous~\cite{canadian2006technical} and current~\cite{canadian2018document} Canadian regulatory documents. For~the independent test, 50 ``thick'' TLD chips were submitted to an external organization where chips were exposed to doses that were not revealed to CNL. The~doses were measured afterward, in-house, by the CNL Dosimetry Services personnel following the standard operational practice. The~results were submitted to the external organization for validation. For~the time when the ``thick'' TLD chips had been deployed in Cube Hall, one could estimate  uncertainty in $ RCF $ as a relative standard deviation of the measurements of 1.6\%~\cite{ACMR-2018}. For~the time when the ``thick'' TLD chips had been removed from Cube Hall and analyzed, one could estimate  uncertainty in $ RCF $ of 2.0\%~\cite{ACMR-2019}. {{The}  organization used for the independent test delivers multiple exposures for ``thick'' TLD chips expressed in milliroentgen with combined uncertainty at the point of measurements of approximately 0.6\%, corresponding to one standard deviation. At~CNL, the calibration process for ``thick'' TLD chips used in this work is linked to external doses expressed in milliroentgen. Therefore, we report ambient dose and dose rate measurements for SNOLAB in roentgen. The~conversion factor from roentgen to SI unit gray will be provided further on.

\color{black}The calibration factor for dosimeters is called the element correction factor, or~$ECF$. The~$ECF$ is used as a multiplier with the reader output to make the response of each dosimeter comparable to the average response of a designated group of dosimeters maintained as calibration dosimeters. %As calibration source for calibration dosimeters we use as irradiation source the natural uranium slab. All tests were performed using natural uranium irradiator at CRL that demonstrates traceability to the National Research Council Canada (NRCC) reference calibration center, the~primary calibration laboratory.
Again, to~compensate for the element response of the TLD, an~$ECF$ is usually used. The~$ECF$ is defined as the light output of one TL element relative to the average light output among similar elements within a group of reference TLDs. {{A} set of calibration dosimeters is used as group reference TLDs. To~generate $ECF$s, a~method for normalizing the light output of an element to the actual delivered dose is usually~used. 

For the determination of $ ECF $ for each ``thick'' TLD chip used in this work, CRL Dosimetry Services also follow the best industrial practices and requirements outlined in the current~\cite{canadian2018document} and previous~\cite{canadian2006technical} Canadian regulatory documents. Moreover, it has been shown that for the TLD chips of the same type as TLD-100~\cite{tawil1996system}, the~$ ECF $ is portable to a different dose. In~other words, the~$ ECF $ obtained for a particular dose could be used for a different dose as~well.

To test the reproducibility, a~``thick'' TLD chip was cleared, exposed to a certain dose, and~read. This cycle was repeated a few times. {The} details on the particular dose used for the $ ECF $ calculation include information owned and developed by CNL that has particular value to CNL’s business interests. The~final result is expressed as a relative standard deviation of the measurements and is less than 3\%. Typical batch homogeneity (or uniformity) for ``thick'' chips is expressed as a relative standard deviation of 4\% to~5\% and represents the variation in examined samples containing a set of TLD-100 chips that are irradiated to a certain dose value. The~uniformity and reproducibility of TLD-100 dosimeters are also studied in detail elsewhere~\cite{miljanic2002main} and reported as 4\% for~both.

\color{black}Prior to $ECF$ generation, the~TLD reader must be calibrated to obtain the correct light output of the TL element. The~$ECF$s for the first irradiation of the reference TLDs are calculated by the following equation: 

\begin{linenomath*}
\begin{equation}
	{ECF_i} = \frac{ A } { R_i},
	\label{Eq:app:ECF}
\end{equation}
\end{linenomath*}
where $ECF_i$ is the element correction factor in detector $i$, $A$ is the average of the counts of the set of calibration dosimeters, and~$R_i$ is the count in the detector $i$ exposed to the same dose as a set of calibration dosimeters. Combining Equations~(\ref{Eq:app:RCF}) and (\ref{Eq:app:ECF}), the~ambient dose could be~obtained. 

In addition to the dosimeters deployed at the water shielding of the DEAP-3600 detector, a~set of representative control dosimeters of the same type was kept close to the DEAP-3600 detector in a well-shielded location during the measuring period. We used a tungsten cube box (with sides approximately 10~cm long and  around 12~mm thick) as shielding to store control dosimeters. The~control dosimeters are used to monitor dose that does not belong to the test exposure of the environmental gamma radiation, that is, the dose to which dosimeters are also exposed while they are not in use (during storage and transport when control dosimeters have been taken out of the shielding). This includes doses from background radiation and other sources that are not related to  testing~exposure. 

When processed, the~average dose determined using the set of control dosimeters is subtracted from the doses determined using deployed dosimeters so that the net ambient dose can be accurately determined. As~we mentioned previously, control dosimeters were shipped with deployed dosimeters during transport in order to identify potential exposure to radiation during transportation and storage. Precautions were taken to protect deployed and control dosimeters from being contaminated with radioactive material. In~order to obtain reliable and valid dose results from the measurement at SNOLAB, a~rigorous quality assurance and quality control program in accordance with Canadian regulatory requirements~\cite{canadian2006technical,canadian2018document} was followed by the CRL Dosimetry Services licensed provider.  The~precision, accuracy, and~reliability of dose estimates by the CRL Dosimetry Services are  successfully checked annually via an independent blind test by an external organization, which is mandated by the Canadian Nuclear Safety Commission~\cite{canadian2018document}.

%%%%%%%%%%%%%%%%%%%%%%%%%%%%%%%%%%%%%%%%%%
\begin{figure}[H]
	\centerline{
		\centering
		\includegraphics[width=0.9 \columnwidth ]{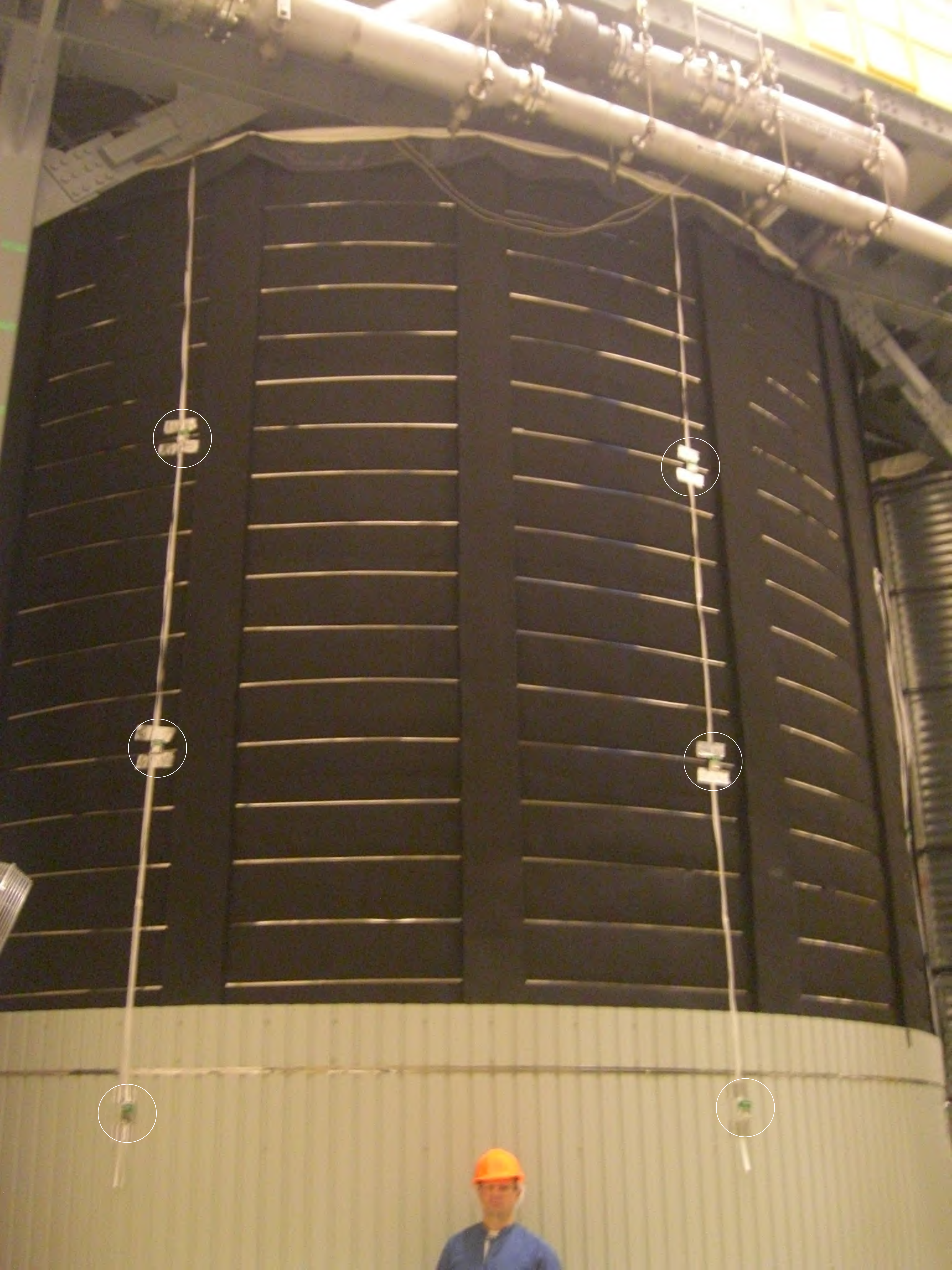} 
	}	
	\caption{Integrating passive TLD detectors deployed at DEAP-3600 water shielding. Six of the badges, circled in the picture, were distributed around the water shield. Each badge contained two TLD-100 chips (see Figure~\ref{fig:badge}) practically at the same spot. TLDs that were hosted inside badge ID 29 (see Table~\ref{tab:res}) were placed at the top of the DEAP-3600 water shielding. The MiniCLEAN water tank is also partially~shown. }
	\label{fig:shield}
\end{figure}

It had been anticipated that ethe nvironmental gamma background at Cube Hall in SNOLAB would be low and that the ambient dose measurements with integrating passive detectors would be close to the minimum detectable level of the TLD technology (in other words, the~signal level would be comparable to the background level)~\cite{currie1968limits}. Therefore,  special attention was applied to improving the quality of the measurements. In~practice, we used two TLD-100 chips in the same badge casing when they were deployed for ambient dose measurements. This option allowed us to check the quality of the ambient dose measurements. If~the ambient dose values at one deployed location were different by more than one standard deviation, these values were not used in the~analysis. {Dose} response linearity and practical factors influencing minimum detectable dose for various thermoluminescent detector types are discussed elsewhere~\cite{harvey2015dose}.

\begin{table}[t] 
	\caption{The measurement results from integrating passive detectors, such as TLDs, for ultra-low-level ambient dose and dose rate at the surface of water shielding at the DEAP-3600 detector (except TLDs with badge IDs 25 and 26, which were placed next to the fire door; see Figure~\ref{fig:CubeHall}).  TLD detectors that were inside badge ID 28 were placed on the deck of Cube Hall. TLD detectors located inside badge~29 were placed on  top of the water shielding of the DEAP-3600~detector. \label{tab:res}}
	\newcolumntype{C}{>{\centering\arraybackslash}X}
	\begin{tabularx}{\textwidth}{CCCCC}
		\toprule
		%\textbf{Title 1}	& \textbf{Title 2}	& \textbf{Title 3}\\
		\begin{tabular}[c]{@{}c@{}}\textbf{Badge}   \\      \textbf{ID}\end{tabular} &
		\begin{tabular}[c]{@{}c@{}}\textbf{Rear} \\      \textbf{Exposure} \\      \textbf{(Milliroentgen)}\end{tabular} &
		\begin{tabular}[c]{@{}c@{}}\textbf{Front} \\      \textbf{Exposure} \\      \textbf{(Milliroentgen)}\end{tabular} &
		\begin{tabular}[c]{@{}c@{}}\textbf{Average} \\      \textbf{Exposure} \\      \textbf{(Milliroentgen)}\end{tabular} &
		\begin{tabular}[c]{@{}c@{}}\textbf{\hl{Average}} \\      \textbf{\hl{Rate}} \\      \boldmath{\textbf{($\upmu$R/h)}}\end{tabular} \\			
		\midrule
%		4  & 4.4 & 5.2  & 4.8~$\pm$~0.6 & 7.7~$\pm$~0.9 \\ %\hline
%		11 & 2.8 & 3.0  & 2.9~$\pm$~0.1 & 4.7~$\pm$~0.2 \\ %\hline
%		12 & 4.1 & 4.1  & 4.1~$\pm$~0.0 & 6.6~$\pm$~0.0 \\ %\hline
%		13 & 2.2 & 2.0  & 2.1~$\pm$~0.1 & 3.4~$\pm$~0.2 \\ %\hline
%		16 & 2.0 & 2.3  & 2.1~$\pm$~0.2 & 3.4~$\pm$~0.3 \\ %\hline
%		17 & 2.5 & 2.7  & 2.6~$\pm$~0.1 & 4.2~$\pm$~0.2 \\ %\hline
%		19 & 5.5 & 4.9  & 5.2~$\pm$~0.4 & 8.4~$\pm$~0.7 \\ %\hline
%		20 & 4.8 & 6.1  & 5.4~$\pm$~0.9 & 8.7~$\pm$~1.5 \\ %\hline
%		23 & 4.1 & 3.7  & 3.9~$\pm$~0.3 & 6.3~$\pm$~0.5 \\ %\hline
%		24 & 5.4 & 5.4  & 5.4~$\pm$~0.0 & 8.7~$\pm$~0.0 \\ %\hline
%		25 & 8.7 & 10.4 & 9.6~$\pm$~1.2 & 15.4~$\pm$~1.9 \\ %\hline
%		26 & 8.9 & 9.2  & 9.1~$\pm$~0.2 & 14.6~$\pm$~0.3 \\ %\hline
%		28 & 3.4 & 2.9  & 3.2~$\pm$~0.4 & 5.1~$\pm$~0.6	\\ %\hline
%		29 & 4.2 & 3.9  & 4.1~$\pm$~0.2 & 6.6~$\pm$~0.3 \\ %\hline
         4 & 4.4 & 5.2 & 4.8 $\pm$ 0.6 & \hl{ 3.6 $\pm$ 0.4} \\
		11 & 2.8 & 3.0 & 2.9 $\pm$ 0.1 & \hl{ 2.2 $\pm$ 0.1} \\
		12 & 4.1 & 4.1 & 4.1 $\pm$ 0.0 & \hl{ 3.1 $\pm$ 0.0} \\
		13 & 2.2 & 2.0 & 2.1 $\pm$ 0.1 & \hl{ 1.6 $\pm$ 0.1} \\
		16 & 2.0 & 2.3 & 2.1 $\pm$ 0.2 & \hl{ 1.6 $\pm$ 0.2} \\
		17 & 2.5 & 2.7 & 2.6 $\pm$ 0.1 & \hl{ 1.9 $\pm$ 0.1} \\
		19 & 5.5 & 4.9 & 5.2 $\pm$ 0.4 & \hl{ 3.9 $\pm$ 0.3} \\
		20 & 4.8 & 6.1 & 5.4 $\pm$ 0.9 & \hl{ 4.1 $\pm$ 0.7} \\
		23 & 4.1 & 3.7 & 3.9 $\pm$ 0.3 & \hl{ 2.9 $\pm$ 0.2} \\
		24 & 5.4 & 5.4 & 5.4 $\pm$ 0.0 & \hl{ 4.0 $\pm$ 0.0} \\
		25 & 8.7 & 10.4 & 9.6 $\pm$ 1.2 & \hl{ 7.1 $\pm$ 0.9} \\
		26 & 8.9 & 9.2 & 9.1 $\pm$ 0.2 & \hl{ 6.7 $\pm$ 0.2} \\
		28 & 3.4 & 2.9 & 3.1 $\pm$ 0.4 & \hl{ 2.3 $\pm$ 0.3} \\
		29 & 4.2 & 3.9 & 4.0 $\pm$ 0.2 & \hl{ 3.0 $\pm$ 0.2} \\		\bottomrule
	\end{tabularx}
\end{table}
%\end{linenomath*}

\section{Results}

Detectors were deployed in the Cube Hall at SNOLAB. The~majority of the TLDs were located around the DEAP-3600 water shield (see Figure~\ref{fig:shield}). Table~\ref{tab:res} provides ambient dose results for  all detectors that pass the selection rule. Both ``rear'' and ``front'' TLD chips (see Figure~\ref{fig:badge}) results are provided as well as the average value and standard deviation. Detectors were placed in Cube Hall at SNOLAB on \hl{Thursday, 27 November 2018} around 2~p.m. and were taken out on Tuesday, 22 January 2019 around~noon. 

The total duration of the ultra-low-level ambient dose measurements in Cube Hall was \hl{55 days and 22 h, or~1342}~h. The~deployment and removal of the TLDs took 2~h in total, and~we took this time as an uncertainty in the time interval measurements; in other words, the~deployment time interval of the passive integrating detectors was \hl{1342 $\pm$ 2~h}.

We applied the direct comparison method to calibrate the timer used for time records. For~this, we followed the recommended practical guide ``Stopwatch and Timer Calibrations''~\cite{gust2009stopwatch} from the National Institute of Standards and Technology (NIST). We  verified, following the guidance in~\cite{gust2009stopwatch}, that the timer was within the 0.02\% tolerance and calibrated to better than 0.02\%. Nevertheless, the~accuracy of the time interval and tolerance of the timer used in the measurements of the ambient dose rate were neglected as they were much less than the standard deviation of the ambient dose measurements (see Table~\ref{tab:res}). 

The roentgen (R) is actually a legacy unit defined operationally as the charge produced by X-rays or gamma rays in one kilogram of air at standard temperature and pressure conditions (1~R = $2.58\times10^{-4} $ C/kg). The~unit is, therefore, still routinely employed in environments where X-rays and gamma rays are found~\cite{cerrito2017radiation}, such as in the Cube Hall cavity at SNOLAB. Values are often given in milliroentgen (mR). Historically, the~roentgen was a convenient unit because early work with radiation was concerned primarily with X-rays, and~radiation was commonly detected with instruments that measured the amount of ionization produced in air. The~units of  dose are the gray (Gy) and the sievert (Sv) in the SI system. In~this work, we would recommend the use of roentgen for measured ambient dose in Cube Hall. The~situation for dose units is similar to that for other SI units competing with older units. The~older units nonetheless remain in common usage in Canada. If~the SI dose unit Gy is required, the conversion factor is 8.8~$\upmu$Gy/mR~\cite{kumar2019radiation,grupen2010introduction}.

\color{black}Since experimental data on ambient dose in Cube Hall come from many locations (most of them around the DEAP-3600 water shielding), with~each having its own characteristic variability, the~problem centers on the appropriate selection of the data to obtain the ``true'' ambient dose value(s) for TLDs occupying the same badge. In~order to achieve this, the~statistical analysis should recognize the existence of variability both within the group (i.e.,~the ambient dose data from TLD badges with the same ID) and between groups (i.e.,~the ambient dose data from TLD badges with different IDs). Both types of variability are considered here to be random effects and are described by their associated components of variance. In~reality, for~a TLD badge with a unique ID, only two TLD chips located at the same place were deployed at Cube Hall. Therefore, we have checked via a calibration process (i.e.,~exposure of a TLD badge with a unique ID to a known dose) that, indeed, the measured dose values coming from the same badge TLD chips are the same within their variance (i.e.,~this is a ``selection criterion'' for ambient dose).

Table~\ref{tab:res} shows solely the results for the TLDs that passed the selection criterion that was based on the comparison of the ambient dose results for the rear and front TLD chips located in the same badge. As~ mentioned previously, it was anticipated and verified during the calibration process that ambient dose values were within one standard deviation for the two TLD chips from the same badge. The~subtracted control dosimeters dose that identified exposure to radiation during transportation and storage was 9.5~$\pm$~1.1~milliroentgen.

The TLD detectors that were inside badges with IDs, from 4 to 24 (see Table~\ref{tab:res}) were placed on the water shielding of the DEAP-3600 detector (see Figure~\ref{fig:shield}). Badge IDs 25 and 26 were placed next to the fire door (see Figure~\ref{fig:CubeHall}). % and at the same place where NaI(Tl) detector was taken data (this will be discussed elsewhere). 
Badge ID 28 was placed on the deck of Cube Hall. Badge ID 29 was placed on the top of the water shield of the DEAP-3600 detector. The~exact locations (i.e.,~within centimeter accuracy) of TLD badges with various IDs on the water shielding of the DEAP-3600 detector and around Cube Hall are not provided in this paper as such a determination was outside the scope of work for these studies (see Section~\ref{sow}).  

%\begin{linenomath*}

%%%%%%%%%%%%%%%%%%%%%%%%%%%%%%%%%%%%%%%%%%
\section{Discussion}
\label{sec:Dis}

As we mentioned  in Section~\ref{sow}, the~first goal for this research was to identify if it is  possible to use integrating passive detectors, such as TLDs, for ambient dose measurements at the SNOLAB low-level background laboratory. As~one can see from Table~\ref{tab:res}, it is indeed possible. The~second goal was to check the uniformity of the background gamma radiation around the water shielding for DEAP-3600. The~ background radiation  around the detector shielding was found to be not uniform.  \color{black}Table~\ref{tab:res} shows that within one standard deviation of the results from TLD dosimeters placed on the water shielding of the DEAP-3600 detector, there is significant non-uniformity in the results (by at least a factor of two). One of the reasons for the measured non-uniformity of environmental gammas around the DEAP-3600 water tank is the existence of the MiniCLEAN water tank in the Cube Hall next to the DEAP-3600 detector (see Figure~\ref{fig:shield}). This result has high importance in the  assessment of background for the DEAP-3600 detector using Monte Carlo simulations. 

On the other hand, \color{black}the ambient dose and dose rate data presented in Table~\ref{tab:res} could be used as  conservative and independent estimates for the upper limit of environmental gammas around the DEAP-3600 water shield. In~other words, one could use the upper limit ambient dose value to estimate the thickness of the shield required for a detector, such as DEAP-3600, to become insensitive to the environmental gamma background. For~that, one should use the ambient dose data only from the integrating passive detectors placed around the DEAP-3600 water shielding. In~other words, the~dose data from Table~\ref{tab:res} for badge IDs from 4 to 24 and badge ID 29 result in the ambient dose for the DEAP-3600 water shielding ($ D_{\rm w.sh.} $) and, taking into account the exposure period of \hl{1342~h}, the~ambient dose rate ($ R_{\rm w.sh.} $) as follows:
\begin{linenomath*}
\begin{equation}
		D_{\rm w.sh.} = 3.9  \pm  1.3 \, {\rm mR \, and~\, } \color{red} R_{\rm w.sh.} = 2.8 \pm 1.0 \, \upmu{\rm R/h}. 
		\label{Eq:D_wsh}
	\end{equation}
\end{linenomath*} \color{black}

Using a similar approach, one could estimate the upper limit of environmental gammas in the Cube Hall for any existing or potential large-scale detector. For~that, one should use all data from Table~\ref{tab:res}, which results in Cube Hall ambient dose ($ D_{\rm CH} $) and ambient dose rate ($ R_{\rm CH} $) as follows:
\begin{linenomath*}
\begin{equation}
		D_{\rm CH} = 4.6  \pm  2.3 \, {\rm mR \, and~\, } \color{red} R_{\rm CH} = 3.4 \pm 1.7 \, \upmu{\rm R/h}.
		\label{Eq:D_CH}
	\end{equation}
\end{linenomath*} \color{black}

The results shown in Equations~(\ref{Eq:D_wsh}) and~(\ref{Eq:D_CH}) are essentially average values and their statistical uncertainties. This approach should be enough for estimates of a conservative upper limit for environmental gammas. An~alternative method that  takes into account weighting factors or algorithms for the most frequent value could also be applied (see details in~\cite{paule1982consensus} and~\cite{steiner1988most,steiner1997optimum}, respectively) but are not discussed in this paper. As~we mentioned previously, Equations~(\ref{Eq:D_wsh}) and~(\ref{Eq:D_CH}) show conservative estimates for the upper limit of environmental gammas around the DEAP-3600 water shield and in the Cube Hall, respectively. Those values are useful for the estimation of the appropriate shielding that would significantly reduce the contribution from ambient dose in the signal search regions. It is obvious from the ambient dose and ambient dose rate results shown in Table~\ref{tab:res} that the distribution of environmental gammas around the DEAP-3600 water shielding is non-uniform. Consequently, due to the limited time of measurements (exposure period was \hl{1342} h), the~observed statistical uncertainty for some ``thick'' TLD chips presented in Table~\ref{tab:res} is large (relative statistical uncertainty of 17\% for badge ID~20). The~statistical uncertainty for ultra-low-level ambient dose and dose rate measurements observed in SNOLAB with ``thick'' TLD chips could be improved if the exposure period was extended to roughly \hl{six} months. However, for~practical operational reasons, the opportunities for such an exposure period are limited but potentially could be available at the end of the DEAP-3600 operation.

\color{black}On the other hand, the~deployment of multiple integrating passive detectors at the same time  reduces the total exposure time that otherwise would be needed with a single active detector sequentially deployed  at multiple locations. For~example, for~the ambient dose data presented in Table~\ref{tab:res}, \hl{1342 $ \times $ 14 $ \times $ 2 = 37,576~h (i.e.,~$ \simeq $1566~days, or~$ \simeq $4~years}) would be required with a single detector placed sequentially at the same locations in Cube Hall.

Direct dark matter search experiments, such as DEAP-3600, are very sensitive to the background in the observation area of the experiment. Typically,  estimates of the background contributions from environmental gamma sources are based on  measurements with thallium-activated sodium iodine (NaI(Tl)) or HPGe detectors that are exposed for a long time at various locations in the underground lab. Examples are,  for~the SNO experiment, a~15.6~kg NaI(Tl) crystal with an exposure of 1490~h (approximately 62~days and 2~h) and a 2.1 kg HPGe detector with an exposure of 20.65~h in the SNO cavern.  A~NaI(Tl) detector,  also located at SNOLAB (see Figure~\ref{fig:SNOLAB3D}), was used to measure the gamma flux~\cite{isaac1997high} up to 10~MeV, while an HPGe detector was used previously to measure the gamma flux up to 6~MeV. {Active} detectors, such as NaI(Tl) and HPGe, for environmental gammas, have been used outside the Cube Hall at SNOLAB~\cite{isaac1997high}, but~those results cannot be compared with ours. However, the~NaI(Tl) active detector was measuring ultra-low-level Cube Hall background at the time that ``thick'' TLD chips were deployed at Cube Hall, and~the results will be available soon.

The~use of a scintillator or semiconductor detector technologies to detect high-energy environmental gammas has a major disadvantage as their efficiency decreases with the increasing energy of the gamma radiation. %\footnote{It would have been interesting to compare the ambient gamma background measurements using HPGe, NaI(Tl), and~TLD technologies at SNOWLAB; however, that was outside the scope of work (see Section~\ref{sow}) for these studies.} 
By contrast, the~use of integrating passive detectors, such as TLD, provides less energy-dependent exposure measurement results for gammas with higher energy~\cite{GUIMARAES2003127,FADZIL2022110232}. Moreover, because~of their size and weight, the~integrating passive detectors, such as TLDs, could be deployed directly at the location where one would like to find the ambient dose from environmental gamma flux. Further, because~those detectors are passive, they could be exposed for a much longer time to improve the accuracy of the ambient dose results. For~example, some environmental TLDs used for measuring ambient dose and ambient dose rates at CRL locations are exposed for 6 or 12~months in the field.

On the other hand, one of the disadvantages of the use of integrating passive detectors, such as TLDs, in measuring ambient dose around an operational dark matter detector, such as DEAP-3600, is their inability to measure a time evolution. In~other words, for~measuring environmental ambient dose from gamma flux, one should not bring any strong external gamma sources that would be used, for~example, to~calibrate the gamma response of a large dark matter detector. Therefore, one of the best times to characterize the environmental ambient dose rate from gamma flux would be a long period before the deployment of a large-scale experiment in the underground location. That would allow the use of the measured ambient dose data for environmental gamma flux as input for Monte Carlo simulations in order to assess the performance of a particular large-scale detector at a certain underground location. 

In principle, the~use of integrating passive detectors, such as TLDs, to measure the ambient dose and dose rate could be complementary to the use of active detector technology, such as a scintillator or semiconductor. The deployment of inexpensive passive technology that provides ambient dose data for the same exposure interval from many locations and that is not dependent on the photon energy up to 10~MeV could be complementary to active detector technology that provides spectral photon information through a  single long measurement~\cite{isaac1997high}. 

Moreover, the~recent successful development of integrating passive technology, such as the optically stimulated luminescence (OSL) technique, provides access to lower-level detection limits~\cite{sommer2006investigation} for detectable activity compared with the TLD technique~\cite{yukihara2014state}. OSL is a functionally straightforward but, at the same time, accurate and
flexible detector technique, providing significant performance and operational advantages in comparison to the related TL technique with its major re-read capability~\cite{yukihara2014state}. Potential applications of the OSL technique for ambient dose and dose rate measurements at a low-level background laboratory, such as SNOLAB, could reduce exposure time while reaching the same level of sensitivity as the TLD technique presented in this paper. 

%%%%%%%%%%%%%%%%%%%%%%%%%%%%%%%%%%%%%%%%%%
\section{Conclusions}

It has been shown that an integrating passive detector, such as the TLD, is very useful for environmental ambient dose and dose rate measurements. It allows the estimation of the environmental gamma background around large-scale experiments, such as the operational DEAP-3600, and~allows the use of those data as input for Monte Carlo simulations to estimate the potential background in regions of interest in the search for dark matter particle candidates, such as WIMPs. It has been shown (see Table~\ref{tab:res}) that the environmental gamma background around DEAP-3600 detector shielding is non-uniform. \color{black}In addition, backgrounds of all types for large-scale dark matter detectors, such as DEAP-3600, must not only be minimized but~ also be measured with robust techniques in~situ. Integrating passive detectors, such as TLDs, are inexpensive, compact, and~easy to manipulate and, therefore, potentially could be used to measure and monitor ambient environmental gamma dose and dose rate in~situ.

Moreover, to~the best of our knowledge, this is the first implementation of integrating passive detectors, such as TLDs, in one of the deepest operational underground low-level background laboratories for ambient dose and dose rate measurements. 

%%%%%%%%%%%%%%%%%%%%%%%%%%%%%%%%%%%%%%%%%%
%\section{Patents}

%This section is not mandatory, but may be added if there are patents resulting from the work reported in this manuscript.

%%%%%%%%%%%%%%%%%%%%%%%%%%%%%%%%%%%%%%%%%%
\vspace{6pt} 

%%%%%%%%%%%%%%%%%%%%%%%%%%%%%%%%%%%%%%%%%%
%% optional
%\supplementary{The following supporting information can be downloaded at:  \linksupplementary{s1}, Figure S1: title; Table S1: title; Video S1: title.}

% Only for the journal Methods and Protocols:
% If you wish to submit a video article, please do so with any other supplementary material.
% \supplementary{The following supporting information can be downloaded at: \linksupplementary{s1}, Figure S1: title; Table S1: title; Video S1: title. A supporting video article is available at doi: link.}

%%%%%%%%%%%%%%%%%%%%%%%%%%%%%%%%%%%%%%%%%%
\authorcontributions{Conceptualization, V.V.G. and O.K.; methodology,  V.V.G.; software, V.V.G.; validation,  V.V.G., J.S., and~O.K.; formal analysis, V.V.G. and O.K.; investigation,  V.V.G. and O.K.; resources,  V.V.G. and O.K.; data curation,  V.V.G. and O.K.; writing---original draft preparation,  V.V.G. and O.K.; writing---review and editing,  V.V.G., O.K., J.S., C.J.J., P.G., and~E.V.-J.; visualization,  V.V.G. and O.K.; project administration,  V.V.G. and O.K.; funding acquisition,  V.V.G. and O.K. All authors have read and agreed to the published version of the~manuscript.}

\funding{This research received no external~funding.}

\institutionalreview{Not applicable.}

\informedconsent{{Not applicable.}} 

\dataavailability{Not applicable.} 

\acknowledgments{{We} owe a debt of gratitude to Cindy Hamel, Alex Angelkovski, and~Kara Stace of CNL, for~their help with the project, preparation of integrating passive detectors, TLD deployment at SNOLAB, and~all organizational steps needed to carry out this project. We also thank the many CNL and SNOLAB personnel, too numerous to mention by name, for~their help. The~authors wish to thank Helena Rummens for her careful editing of the paper. We thank the Natural Sciences and Engineering Research Council of Canada, the~Canadian Foundation for Innovation, the~Ontario Ministry of Research and Innovation, and~Alberta Advanced Education and Technology (ASRIP). 
We thank SNOLAB and its staff for support through underground space and logistical and technical services. SNOLAB operations are supported by the Canada Foundation for Innovation and the Province of Ontario Ministry of Research and Innovation, with~underground access provided by Vale at the Creighton mine site. We thank Vale for their continuing~support.
}

\conflictsofinterest{The authors declare no conflict of~interest.} 

%%%%%%%%%%%%%%%%%%%%%%%%%%%%%%%%%%%%%%%%%%
%% Optional
%\sampleavailability{Samples of the compounds ... are available from the authors.}

%% Only for journal Encyclopedia
%\entrylink{The Link to this entry published on the encyclopedia platform.}

\abbreviations{Abbreviations}{
The following abbreviations are used in this manuscript:\\

\noindent 
\begin{tabular}{@{}ll}
CNL & Canadian Nuclear Laboratories\\ 
%CNSC& Canadian Nuclear Safety Commission\\
CRL & Chalk River Laboratories\\
CPE & charge-particle equilibrium\\
DEAP& Dark matter Experiment using Argon Pulseshape discrimination\\
ECF & element correction factor\\
HPGe& high-purity germanium \\ 
ID  & identification\\ 
LLBL& Low-Level Background Laboratory\\ 
NEWS-G& New Experiments With Spheres-Gas\\ 
%NIST& National Institute of Standards and Technology\\ 
TL  & thermoluminescent\\
TLD & thermoluminescent dosimeter\\
RCF & reader calibration factor\\ 
%PET & Polyethylene Terephthalate\\  
PMT & photomultiplier tube\\ 
%PVC & Polyvinyl Chloride\\ 
OSL & optically stimulated luminescence\\
SNO & Sudbury Neutrino Observatory \\
SNOLAB & a lab created to host other neutrino and dark \\ 
    & matter experiments after the success of SNO experiment \\
WIMP & weakly interacting massive particles 
\end{tabular}
}

%%%%%%%%%%%%%%%%%%%%%%%%%%%%%%%%%%%%%%%%%%
%% Optional
%\appendixtitles{no} % Leave argument "no" if all appendix headings stay EMPTY (then no dot is printed after "Appendix A"). If the appendix sections contain a heading then change the argument to "yes".
%\appendixstart
%\appendix
%\section[\appendixname~\thesection]{}
%\subsection[\appendixname~\thesubsection]{}
%The appendix is an optional section that can contain details and data supplemental to the main text---for example, explanations of experimental details that would disrupt the flow of the main text but nonetheless remain crucial to understanding and reproducing the research shown; figures of replicates for experiments of which representative data are shown in the main text can be added here if brief, or as Supplementary Data. Mathematical proofs of results not central to the paper can be added as an appendix.
%
%\begin{table}[H] 
%\caption{This is a table caption.\label{tab5}}
%\newcolumntype{C}{>{\centering\arraybackslash}X}
%\begin{tabularx}{\textwidth}{CCC}
%\toprule
%\textbf{Title 1}	& \textbf{Title 2}	& \textbf{Title 3}\\
%\midrule
%Entry 1		& Data			& Data\\
%Entry 2		& Data			& Data\\
%\bottomrule
%\end{tabularx}
%\end{table}
%
%\section[\appendixname~\thesection]{}
%All appendix sections must be cited in the main text. In the appendices, Figures, Tables, etc. should be labeled, starting with ``A''---e.g., Figure A1, Figure A2, etc.

%%%%%%%%%%%%%%%%%%%%%%%%%%%%%%%%%%%%%%%%%%
\begin{adjustwidth}{-\extralength}{0cm}
%\printendnotes[custom] % Un-comment to print a list of endnotes

\reftitle{References}

% Please provide either the correct journal abbreviation (e.g. according to the “List of Title Word Abbreviations” http://www.issn.org/services/online-services/access-to-the-ltwa/) or the full name of the journal.
% Citations and References in Supplementary files are permitted provided that they also appear in the reference list here. 

%=====================================
% References, variant A: external bibliography
%=====================================
%\bibliography{your_external_BibTeX_file}

%\bibliographystyle{elsarticle-harv} 

%\PublishersNote{}

%%%%%%%%%%%%%%%%%%%%%%%%%%%%%%%%%%%%%%%%%%
%% for journal Sci
%\reviewreports{\\
%Reviewer 1 comments and authors’ response\\
%Reviewer 2 comments and authors’ response\\
%Reviewer 3 comments and authors’ response
%}
%%%%%%%%%%%%%%%%%%%%%%%%%%%%%%%%%%%%%%%%%%
\end{adjustwidth}
\end{document}